\documentstyle[12pt,epsfig]{article}
\def\mathrm#1{\rm{#1}}
\newcommand{\be}{\begin{equation} }
\newcommand{\ee}{\end{equation} }
\def\docnum#1{\hbox to \hsize{\hskip123mm\hbox{#1}\hss}}
\def\date#1{\edef\@temp{#1}\ifx\@temp\@empty\def\@temp{\today}\fi
\hbox to \hsize{\hskip123mm\hbox{\@temp}\hss}}
\def\title#1{\vskip1em\begin{center}%
{\normalsize\bf#1\par}\vskip1.5em\end{center}}
\def\@makefnmark{\hbox{$^{\@thefnmark)}$}}
\def\author#1{%% Treat the list of authors
\setcounter{footnote}{0}\def\@currentlabel{}%
\begingroup\def\thefootnote{\arabic{footnote}}
\def\@makefnmark{\hbox{$^{\rm\@thefnmark)}$}}
\global\@topnum\z@ \begin{center}{\lineskip.5em
\begin{tabular}[t]{c}#1\end{tabular}\par}
\end{center}\par\vskip1.5em\@thanks\endgroup}

\def\abstract{\begin{center}{\bf Abstract}\end{center}\quotation}

\begin{document}

\begin{flushright}
{\bf 16 January, 1997}
\end{flushright}

\centerline{\large\bf Variable Field Bending Magnets for Recirculating Linacs}
%\centerline{\large\bf (Early Draft)}

\vskip 2.0 cm

\centerline{B. J. King}

\vskip 0.5 cm

%\centerline{DESY}
%\centerline{F-OPAL}
%\centerline{Notkestrasse 85}
%\centerline{22607 Hamburg}
%\centerline{Deutschland}
\vskip 0.5 cm
\centerline{email: Bruce.King@cern.ch}
\vskip 2 cm

\centerline{\large\bf Abstract}

\vskip 0.5 cm

    A lattice of single aperture superconducting variable field
bending magnets is proposed as a cheap and practical way to
recirculate the beams in recirculating linear accelerators.
It is shown that the VFBM's can be configured to provide strong
focusing in both transverse planes for the full range of beam
momenta transported by the lattice.

\vskip 0.5 cm

\pagebreak

\section{Introduction}
%%%%%%%%%%%%%%%%%%%%%%

   Recirculating linacs, such as that at CEBAF and those proposed
for future muon colliders~\cite{mufs}, economise on RF cavities by circulating
the beam through them several times, at increasing beam energies.
This scheme transfers a large fraction of the cost of acceleration
to the magnet lattices which bend the beam around to return it
to the RF cavities, so it is important to construct the bending
lattice as cheaply as possible.

   A lattice of conventional superconducting dipole magnets, such as
used in storage rings, cannot be used for recirculating linacs,
since they cannot be ramped quickly enough
to keep up with the increase in beam energy in successive passes
through the arcs. Existing schemes for the lattice
include~\cite{mufs} multiple aperture superconducting magnets,
such as used at CEBAF, and interspersing single aperture
superconducting magnets with fast ramping warm magnets.
However, both of these options are expected to be relatively
expensive compared to single aperture superconducting magnets.

   This note introduces the idea of a lattice of single aperture
superconducting
variable field bending magnets (VFBM's). The point of using a
variable field is that successive passes of the beam, at increasing
beam energies, can be deflected by equal amounts by steering the
beam to progressively higher field regions of the apertures.
It will be shown that such magnets can be arranged in a bending
lattice which is strongly focusing in both transverse views.

   A local coordinate frame for the magnets will be used such that
the beam travels in the z direction, the bend direction is horizontal
and along the x direction, and the y coordinate gives the
vertical displacement.

   Throughout the paper, small angle approximations will be used for
the deflection of the beam in each magnet. Also,
it will be assumed that each magnet has a constant field along the
z direction and end effects due to the finite lengths of the magnets
will be neglected. These approximations should not affect the general
validity of the concept. Unless otherwise specified, magnetic fields
are given in units of Tesla, lengths in meters, currents in Amperes
and beam momenta in units of GeV/c.

   The note is organised as follows.
The following section gives a general discription of the magnetic
fields in VFBM's and their focusing properties, and introduces the
concept of a strong focusing lattice of VFBM's with alternating
gradients, in close analogy to the conventional strong focusing
lattice using quadrupole magnets. It is noted that the strong
focusing property will apply to all momenta if the magnet focal
length can be made independent of beam momentum. Section 3 gives
an explicit prescription for doing this, by an appropriate
choice of beam trajectories and field distributions. Section
4 addresses the question of how to design the coil configuration
to produce the desired VFBM fields and section 5 provides an
illustrative set of values for the magnet and lattice parameters.
Further studies that will be needed to further assess the idea of
a VFBM lattice are outlined in section 6, before summarizing the
note in the final section.

\section{Overview of Magnets and Lattice}
%%%%%%%%%%%%%%%%%%%%%%%%%%%%%%%%%%%%%%%%%

   The VFBM's are assumed to have a field which is independent of
the coordinate along the beam,
\begin{equation}
\vec{B} \equiv \vec{B}(x,y),
\end{equation}
and there is no field component along the
beam:
\begin{equation}
B_z \equiv 0.
\end{equation}

   For all beam energies, the beam center will always be assumed to
pass through the x-axis, i.e. y=0, which is a symmetry axis for the
magnetic field and which will be referred to as the ``beam plane''.
This implies that the field is vertical in the bend plane,
\begin{equation}
\vec{B}(x,y=0) \equiv B_y(x) \hat{y},
\end{equation}
and the horizontal component of the field is identically zero:
\begin{equation}
B_x(x,y=0) \equiv 0.
                        \label{eq:zerob}
\end{equation}

    A VFBM of length $l$ will bend
a beam of momentum p and position $x$ through an angle
\begin{equation}
\theta(x,p) = \frac{0.3B_y(x)l}{p}. \label{eq:theta}
\end{equation}
In addition, since the field gradient along
the x axis is non-zero
the beam will be either focused or defocused in the
bend plane with a focal length,
$f$, given by:
\begin{equation}
f = \pm \frac{p}{0.3Gl},  \label{eq:focal}
\end{equation}
where
\begin{equation}
G \equiv \frac{\partial B_y}{\partial x}.    \label{eq:gdef}
\end{equation}

    The plus and minus signs in equation~\ref{eq:focal} imply that
the beam is focused or defocused in the bend plane, respectively.
Which of the two cases applies depends
on the beam charge sign and whether it is
travelling in the positive or negative z direction. For example, if
the choice is made such that the bend direction is in the positive
x direction then it is clear that a positive field gradient
is defocusing (the field is stronger on the inside of the bend)
and a negative gradient is focusing.

    The gradients of the magnetic field in the beam plane (y=0) are
constrained by Maxwell's equations in vacuo.
The vanishing of the divergence of $B$ implies that
\begin{equation}
\frac{\partial B_y}{\partial y} = -\frac{\partial B_x}{\partial x}.
\end{equation}
This is identically zero in the bend plane, from equation~\ref{eq:zerob},
which simply means that the bending field remains constant to first order
above and below the bend plane.

    Since the field is independent of z, the x- and y-components of
Maxwell's curl equation are trivially zero. More important is the
z-component of the curl equation:
\begin{equation}
\frac{\partial B_x}{\partial y} = \frac{\partial B_y}{\partial x}.
\end{equation}
This is easily seen to imply that the focal length in the vertical
(y) direction will have equal magnitude and opposite sign to that in
the horizontal (x) direction. In other words, if the magnet is focusing
with a given strength in the x direction then it will be defocusing
with equal strength in the y direction, and vice versa.

   Of course, this focusing property is exactly the same as in a
quadrupole magnet, so a lattice of VFBM's will have exactly the same
focusing properties as a quadrupole lattice with the same magnet focal
lengths. In particular, it is obvious that a lattice
of VFBM's with equal magnetic fields and alternating gradients at the
aperture center, (x=0, y=0), can be arranged to produce strong focusing
in both planes for a beam of some chosen momentum passing through the
aperture centers, exactly as is normally done with quadrupole magnets.
We will refer to this chosen momentum as the central momentum, $p_0$.

   The main point of this paper is to demonstrate that the strong
focusing behaviour at the central momentum can be applied identically
to all the beam momenta passing through the magnet lattice. This is
achieved by arranging the beam trajectories and variation of the
on-axis magnetic field such that the focal lengths for all momenta
are equal to those at $p_0$, thus ensuring that the focusing properties
are identical. An explicit prescription for doing this forms the
topic of the following section.

\section{Calculating Beam Trajectories and On-Axis Magnetic Fields}
%%%%%%%%%%%%%%%%%%%%%%%%%%%%%%%%%%%%%%%%%%%%%%%%%%%%%%%%%%%%%%%%%%%

   The goal of this section is to find the magnetic field variation
in the VFBM's and the beam positions, $x^{\pm}(p)$,
which will give the correct bend angle in the lattice along
with optimal strong focusing of the beam in both views. The
superscript ``$\pm$'' in the $x^{\pm}(p)$ distinguishes between
the two types of VFBM's used in the lattice: the ``plus'' refers
to the magnets with the positive on-axis
field gradient in the positive x direction and the ``minus'' to those
magnets with negative field gradient in the x-view.

   For definiteness,
we assume the bend direction to be towards the positive x-axis,
in which case the plus magnets are defocusing in the
bend plane and the minus magnets are focusing, as already mentioned.

   The beam position can be solved order-by-order by
expanding in a Taylor series about the central momentum, $p_0$.
In this section we will only solve for the first order terms:
\begin{equation}
x^{\pm}(p) = K^{\pm}.(p-p_0) +... ,
                             \label{eq:xexpand}
\end{equation}
where $K^+$ and $K^-$ are constants for the plus and minus magnets,
respectively, and both zeroth order terms vanish due to the
definition of $p_0$ and the choice of coordinate system:
\begin{equation}
x^{\pm}(p_0) \equiv 0.
\end{equation}

   The field-times-length of each of the
magnets is conveniently defined for MKSA units as:
\begin{equation}
b(x) \equiv 0.3 B_y(x,y=0)l.
\end{equation}
This definition is chosen so the bend angle of the magnet,
equation~\ref{eq:theta}, takes the simple form:
\begin{equation}
\theta (x,p) = \frac{b(x)}{p}.
\end{equation}

   The field-times-length
can also be expanded in a Taylor series:
\begin{equation}
b^{\pm}(x) = b_0 \pm g_0.x \pm g^{\pm}_1.\frac{x^2}{2} +... .
                        \label{eq:bexpand}
\end{equation}
In this equation the $x$ values are understood to be those of the
beams at the given momentum, $x=x^{\pm}(p)$, and we have used
the assumptions that the bending fields at the central momentum,
$p_0$, are equal and the field gradients are equal but with opposite
signs. The derivative of equation~\ref{eq:bexpand} gives the
Taylor expansions for the field gradients:
\begin{equation}
\frac{\partial b^{\pm}(x)}{\partial x} \equiv g^{\pm}(x)
                              = \pm g_0 \pm g^{\pm}_1.x +...
                        \label{eq:gexpand}
\end{equation}
It is seen that equation~\ref{eq:focal}, for the magnet focal
lengths in the bend plane, can be written as:
\begin{equation}
f^{\pm} = \mp \frac{p}{g^{\pm}}.
                        \label{eq:fpm}
\end{equation}

    Since strong focusing would be expected to be most effective
for focal lengths roughly equal to the spacing between magnet centers,
$L$, it simplifies the equations (but is not necessary) to assume
that this condition holds exactly, i.e.:
\begin{equation}
f^{\pm} \equiv L \equiv \frac{p_0}{g_0}.
                         \label{eq:strfoc}
\end{equation}
These equivalences use equation~\ref{eq:fpm} and
the key assumption that the
focal length can be chosen to be independent of beam momentum.

    We are now ready to derive the displacement constants $K^+$
and $K^-$. For non-zero $x^{\pm}$ the bend angles in the plus and
minus magnets, $\theta^{\pm}(p)$ won't be equal to the central
bend angle,
\begin{equation}
\theta_0 = \frac{b_0}{p_0}.
\end{equation}
Instead, they are easily seen to be given by:
\begin{equation}
\theta^{\pm}(p) = \theta_0 \pm 2 \frac{x^+(p)-x^-(p)}{L}.
                        \label{eq:thetapm}
\end{equation}
(The average of the two bend angles is equal to
$\theta_0$, as it must be.)

   On expanding these equation~\ref{eq:thetapm} to first order in $x^{\pm}$
and $p-p_0$, using
equations~\ref{eq:xexpand},~\ref{eq:bexpand},~\ref{eq:gexpand},
~\ref{eq:fpm} and~\ref{eq:strfoc},
and solving for $K^+$ and $K^-$
one easily obtains:
\begin{equation}
K^+ = 5 \frac{b_0}{g_0p_0}
\end{equation}
and
\begin{equation}
K^- = 3 \frac{b_0}{g_0p_0}.
\end{equation}
From the definitions of $K^+$ and $K^-$, equation~\ref{eq:xexpand},
this can be rewritten in terms of the displacements:
\begin{equation}
x^+(p) = 5 \frac{b_0}{g_0}\, \frac{p-p_0}{p_0}
                             \label{eq:xpp}
\end{equation}
and
\begin{equation}
x^-(p) = 3 \frac{b_0}{g_0}\,\frac{p-p_0}{p_0}.
                             \label{eq:xmp}
\end{equation}
This is the desired first order approximation to the beam trajectories
for beam momenta different from the central momentum, $p_0$.

    Now that the beam trajectories are known to first order it is
possible to apply the strong focusing assumption of equation~\ref{eq:strfoc}
to obtain the first order change in the field gradient. From
equation~\ref{eq:fpm} we obtain:
\begin{equation}
\frac{p_0}{g_0} = \frac{p}{g_0+g^{\pm}_1.K^{\pm}(p-p_0)+...}.
\end{equation}
Solving these equations to the lowest nontrivial order gives,
after some algebra:
\begin{equation}
g^+_1 = \frac{g^2_0}{5b_0}
\end{equation}
and
\begin{equation}
g^-_1 = \frac{g^2_0}{3b_0}.
\end{equation}

   The first order coefficients in the gradient of the field that have
just been derived are also, of course, the second order coefficients in
the field itself. Since the first order derivation of the displacements
$x^{\pm}(p)$ needed only the first order coefficients in the field it is
clear that the second order field coefficients can be used to derive
the second order correction to the displacements. In turn, the second
order displacement coefficients will permit the derivation of the
third order field coefficients, and so on.

   In summary, alternate applications of the constraints on the
bending fields and on the focal lengths enable the Taylor expansion
coeffients of the magnetic field and the beam positions to be determined
to arbitrarily high orders. Hence, the ideal beam positions for all momenta
and the ideal on-axis magnetic field throughout the magnets can, in
principle, be predicted to arbitrary accuracy.

   Since the strong focusing principle would be expected to work
for a range of focal lengths about the optimal value, albeit less
effectively, our requirement that that focal length takes the
optimal value for all momenta is unnecessarily strict. In practice,
the ``rigorous'' solutions for $x^{\pm}(p)$ and $b^{\pm}(x)$ obtained
using the method outlined in this section could be used as a
starting point for design iterations which might compromise the
strong focusing power of the lattice for some momenta in order to
improve on other features of the magnet design.

\section{Layout of Magnet Coils}
%%%%%%%%%%%%%%%%%%%%%%%%%%%%%%%%

   The preceding section specifies a magnetic field
distribution, $B_y(x,y=0)$, which is
is smooth and monotonically varying but which cannot be
expressed in closed form. Obviously, the current distribution to
produce this field must be obtained by numerical means.
This section describes a general minimization procedure to obtain
a suitable layout for the conducting coils, and illustrates the method
using a simple example.

    In general, the desired magnetic field along the x axis will
be produced by a 2-dimensional current distribution  around
the magnet aperture, $J(x,y)$, which is symmetric under reflection
in the x-axis:
\begin{equation}
J(x,y) \equiv J(x,-y).
\end{equation}
This current distribution will produce a magnetic field on the x-axis
with zero horizontal component, $B_x(y=0) \equiv 0$, and a vertical
component given by:
\begin{equation}
B(x) \equiv B_y(x,y=0)
 = 10^{-7} \int dx'dy'J(x',y') \frac{(x-x') }{ (x-x')^2+y'^2 },
                                      \label{eq:bj}
\end{equation}
using MKSA units.

   The goal is to obtain an appropriate current distribution which
gives an on-axis magnetic field closely approximating the desired
field, $B^{true}(x)$. This can be achieved in the following steps:
\begin{enumerate}
   \item specify the regions which can contain conductor and parameterize
a sensible current distribution in these regions in terms of a small
number of adjustable parameters, $C_i$:
\begin{equation}
J(x,y) = J(x,y;C_i), \;i=1,n.
\end{equation}
   \item define an error function to quantify the deviation of the
on-axis field produced by the current distribution from the desired
field. An appropriate error function is:
\begin{equation}
E[C_i]=\int dx (\frac{B^{true}(x)-B(x)}{B^{true}(x)})^2 / (x_{max}-x_{min}),
                                      \label{eq:erf}
\end{equation}
where the magnetic field $B(x)$ is given by equation~\ref{eq:bj} with the
current distribution specified by the values of the $C_i$.
   \item vary the $C_i$ to minimize the error function.
\end{enumerate}

%   General procedure is to parameterize the regions of conductor
%around the magnet aperture and then do a minimization procedure
%to find the current densities to produce the desired magnetic
%field map.
%   This section illustrates the procedure by presenting an
%explicit calculation for a simplistic case:

    To illustrate and test the procedure, an explicit current
distribution was derived for the following simple case:
\begin{itemize}
  \item an exponentially varying magnetic field along the x-axis:
$B(x)=e^x$, for x in the range -1 to 1.
  \item B field from surface current along wedge-shaped magnet aperture.
  \item no requirement that the current sum to zero. This is equivalent
        to assuming that the excess current is returned at a very large
        distance from the aperture.
\end{itemize}

   In more detail, the surface current, $K(x,y(x))$ was parameterized
to have a quadratic form:
\begin{eqnarray*}
K(x,y(x)) & = & C_1 + C_2.x + C_3.(2x^2-4),\;\;\; {\rm for}\; -2<x<2  \\
          & = & 0,\;\;\;\;\;\;\;\;\;\;\; {\rm otherwise}.
\end{eqnarray*}
The two y coordinates of the current for each x, symmetric about the
x axis, were specified by a linear
form with one free parameter and a minimum aperture of 0.1 units at $x=-2$:
\begin{equation}
y(x) = \pm [0.1 + C_4.(x+2)].
\end{equation}

    The MINUIT minimalization software package~\cite{minuit} was
used to find the values of the $C_i$ which minimized the error function
of equation~\ref{eq:erf}.
Numerical integrations were used to evaluate the error function
and the on-axis magnetic field of equation~\ref{eq:bj}. The
constant factor in front of the magnetic field equation
was neglected, corresponding to an overall scale factor in the
magnetic field strength.

    The optimal current distribution was obtained for the parameter
values:
\begin{equation}
C_1=7.228;\;\;\; C_2=2.073;\;\;\; C_3=1.107;\;\;\; C_4=0.656.
\end{equation}
Figure 1 displays the x distribution of this current and figure
2 illustrates the level of agreement between the resulting magnetic
field and the
exponential distribution. The root mean square deviation of the
on-axis magnetic field from the desired exponential
form, given by the square root of the error function, was found
to be 1.9\% for the region between $x=-1$ and $x=1$.

\begin{figure}[htb]
\begin{center}
\mbox{\epsfxsize=12cm\epsffile[60 0 500 650]{current.ps}}
\end{center}
\caption{The surface current distribution used to produce
an approximately exponential bending field in the VFBM.}
\label{current figure}
\end{figure}

\begin{figure}[htb]
\begin{center}
\mbox{\epsfxsize=12cm\epsffile[60 0 500 650]{bfield.ps}}
\end{center}
\caption{The bending field in the VFBM (solid line) produced
by the surface current distribution of figure 1. The dashed
curve is the ``ideal'' exponential field that the current distribution
was tuned to reproduce.}
\label{bfield figure}
\end{figure}

   It is clear that the procedure can be modified to work for a
more realistic magnet design. It is also obvious that infinitely many
conductor configurations can be chosen to produce an acceptably good
approximation to the desired field along the symmetry axis of the magnet.
The decision between possible configurations can therefore be based
on other factors, such as good field quality off-axis, simplicity of
production, a desirable aperture shape, cheap cost and good mechanical
properties.

\section{Example Lattice Parameters}
%%%%%%%%%%%%%%%%%%%%%%%%%%%%%%%%%%%%

   Table 1 gives an illustrative example set of parameters for the VFBM lattice
of the final recirculating linac in a muon collider with a centre of mass
collision energy of about 4 TeV. The values of these parameters should
not be taken too seriously. They have not been optimized or particularly
carefully chosen and their only purpose is to give a rough feel for
the parameter values that might be expected for a more realistic lattice.

   The first 4 parameters in the table,
$p_0$, $B_0$, $G_0$ and $l$, essentially define the lattice
at the central momentum value. The next three parameters,
$f$, $L$ and $R$, follow from relations given in the preceding
sections.

   For a beam with position divergence $<x>$ and angular divergence
$<\phi>$ it is assumed that a strong focusing lattice of focal
length $f$ will have a maximum 1-sigma beam envelope, $S$
of order:
\begin{equation}
S \sim <x> + <\phi>f.
\end{equation}
(And similarly for the y coordinate.)
  If the phase space in the x view,
\begin{equation}
P_x \equiv <x><\phi>,
\end{equation}
is assumed to be fixed independent of the values of the
two terms, $<x>$ and $<\phi>$, then,
\begin{equation}
S \sim <x> + \frac{P_x}{<x>}f
\end{equation}
and this is minimized for
\begin{equation}
<x> \sim  \sqrt{P_x.f},
\end{equation}
at a value of
\begin{equation}
S \sim 2\sqrt{P_x.f}.
\end{equation}
This gives a numerical value of 0.6 mm at 1 TeV and using
the phase space size, $P_x \sim 10^{-8}$ m.rad, of the same order
as assumed in reference \cite{mufs}.

   The maximum and minimum momenta accepted by the lattice have
somewhat arbitrarily been assumed to be factors of two greater
than and less than $p_0$, respectively.
The average bending field needed
for $p_{max}$ is therefore twice as big as that for $p_0$.
Presumably, almost all of the bending power will come from
the ``plus'' magnets, requiring another factor of two stronger
field in these magnets. Hence the maximum field, $B_{max}$ might
be roughly four times larger than $B_0$, and the minimum field
close to zero. The maximum and minimum gradients, $G_{max}$
and $G_{min}$, follow from the central field and gradient by
scaling in proportion to the momentum using equation~\ref{eq:strfoc}.

   The height, $Y_{aperture}$, of the aperture at the central x value,
x=0, was chosen to be about 30 sigma wide at the maximum beam size, $S_0$.
The width of the aperture in $x$, $X_{aperture}$, can be estimated
simply by the dimensional argument of dividing the maximum magnetic
field by the central gradient:
\begin{equation}
X_{aperture} \sim \frac{B_{max}}{G_0}.
\end{equation}

\begin{table}
\centering
\begin{tabular}{|r|c|}
\hline
    parameter                  & value \\
central momentum, $p_0$        &    1 TeV/c  \\
central field, $B_0$           &    1.5 T    \\
central field gradient, $G_0$  &    40 T/m \\
magnet length, $l$             &    8 m     \\
focal length,    $f$           &    10 m     \\
lattice spacing, $L$           &    10 m     \\
bending radius of lattice, $R$ &    2.8 km   \\
maximum beam size at $p_0$, $S$&    ~0.6 mm \\
maximum momentum, $p_{max}$    &    ~2 TeV/c  \\
minimum momentum, $p_{min}$    &    ~0.5 TeV/c \\
maximum field, $B_{max}$       &    ~6 T     \\
minimum field, $B_{min}$       &    ~0 T     \\
maximum gradient, $G_{max}$       & ~80 T/m  \\
minimum gradient, $G_{min}$       & ~20 T/m  \\
aperture height, $Y_{aperture}$ &   ~2 cm  \\
aperture width, $X_{aperture}$ &   ~15 cm  \\
\hline
\end{tabular}
\caption{
{\small \bf
     Example parameters for the VFBM lattice of the final
recirculating linac in a muon collider with a centre of mass
collision energy of about 4 TeV. See text for further details.
}}
\label{tab:example lattice parameters}
\end{table}

%   If it is assumed that the length scale is such that one unit
% corresponds to 10 cm and that the maximum surface current density
% has a realistic value of ???? A/cm then the maximum magnetic field
% in the good field region of the aperture, at $x=1$, is ???? Tesla.

\section{Outlook}
%%%%%%%%%%%%%%%%%

    The following studies still need to be undertaken to confirm that
strong focusing VFBM lattices are feasible and practical for
recirculating linacs:
\begin{enumerate}
   \item continue to higher order the Taylor series expansion of the
on-axis magnetic
field of the VFBM's. This will provide a better estimate of the range of
beam momenta which can be accepted by a VFBM lattice.
   \item use the procedure of section 4 to
determine a realistic and appropriate magnet coil configuration
that will produce the desired on-axis magnetic fields.
   \item perform computer-based ray-tracing simulations of a beam
through a VFBM lattice, to check that it performs as expected.
\end{enumerate}

   If the bending lattice performs as hoped then it will still need to
be matched to the linacs for each pass of the beam, in beam position
and direction and in the phase of the RF cavities. This could possibly
be done using a dispersive section of superconducting magnets or, if
this is found to be impractical, by using fast ramping warm magnets.

\section{Conclusions}
%%%%%%%%%%%%%%%%%%%%%

    The idea of using strongly focusing lattices of VFBM's
in recirculating linear accelerators has been found to be
quite promising and worthy of further study.

\pagebreak

\end{document}